\documentclass[floats,floatfix,showpacs,amssymb,prd,twocolumn,superscriptaddress,nofootinbib]{revtex4-1}

\usepackage{amssymb,amsmath,verbatim,mathtools,needspace,enumitem,etoolbox,graphicx,physics,microtype,afterpage,bm}

\usepackage[dvipsnames, usenames]{xcolor}

\definecolor{linkcolor}{rgb}{0.0,0.3,0.5}
\usepackage[unicode, colorlinks=true, linkcolor=linkcolor, citecolor=linkcolor, filecolor=linkcolor,urlcolor=linkcolor, pdfusetitle]{hyperref}
\usepackage[all]{hypcap}
\usepackage[T1]{fontenc}
\usepackage[utf8]{inputenc}
\usepackage{tabularx}
\usepackage{float}
\allowdisplaybreaks
\graphicspath{{./figure/}}

\newcommand{\jhu}{\affiliation{Department of Physics and Astronomy, Johns Hopkins University, 3400 N. Charles
Street, Baltimore, MD 21218, USA}}
\newcommand{\LIGOlabMIT}{\affiliation{LIGO Laboratory, Massachusetts Institute of Technology, 185 Albany St, Cambridge, MA 02139, USA}}
\newcommand{\MIT}{\affiliation{Department of Physics and Kavli Institute for Astrophysics and Space Research, Massachusetts Institute of Technology, 77 Massachusetts Ave, Cambridge, MA, 02139, USA}}

\definecolor{rb4}{HTML}{27408B}

\begin{document}

\title{Gravitational-wave signal-to-noise interpolation via neural networks}

\author{Kaze W. K. Wong} 
\email{kazewong@jhu.edu}
\jhu

\author{Ken K. Y. Ng}
\email{kenkyng@mit.edu}
\LIGOlabMIT
\MIT

\author{Emanuele Berti}
\email{berti@jhu.edu}
\jhu

\date{\today}

\begin{abstract}
  Computing signal-to-noise ratios (SNRs) is one of the most common tasks in gravitational-wave data analysis.  While a single SNR evaluation is generally fast, computing SNRs for an entire population of merger events could be time consuming.  We compute SNRs for aligned-spin binary black-hole mergers as a function of the (detector-frame) total mass, mass ratio and spin magnitudes using selected waveform models and detector noise curves, then we interpolate the SNRs in this four-dimensional parameter space with a simple neural network (a multilayer perceptron). The trained network can evaluate $10^6$ SNRs on a 4-core CPU within a minute with a median fractional error below $10^{-3}$.  This corresponds to average speed-ups by factors in the range $[120,\,7.5\times10^4]$, depending on the underlying waveform model. Our trained network (and source code) is publicly available online~\cite{NeuralSNR}, and it can be easily adapted to similar multidimensional interpolation problems.
\end{abstract}

\maketitle

\section{Introduction} \label{sec:Introduction}

The paradigm of gravitational wave (GW) astronomy is rapidly shifting from single-source characterization to the understanding of the entire population of merging compact binaries.
The first two observing runs of the LIGO/Virgo Collaboration (LVC) produced a catalog of 10 binary black-hole mergers and one binary neutron star merger~\cite{LIGOScientific:2018mvr}, and other candidates were identified through independent offline analysis of public LVC data~\cite{Nitz:2018imz,Venumadhav:2019lyq,Zackay:2019btq,Nitz:2019hdf}. A total of 67 candidate binary merger events were publicly announced by the end of the third LVC observing run (O3)~\cite{gracedb}, and future upgrades of the detector network are expected to increase the observed merger population by orders of magnitude~\cite{Baibhav:2019gxm}.
Data analysis tools must improve in accuracy and speed in order to keep up with the accelerating pace of GW detections and to capitalize on the growing number of observations.

This work was motivated by two practical considerations. First of all, the simplest test of the viability of astrophysical population models is to compare the predicted GW detection rate against LVC data.
This requires the construction of a synthetic population from the model, and the calculation of SNRs for each event in the synthetic population.
Depending on waveform models and source properties, calculating the SNR of a single event can be time consuming~\cite{Marsat:2018oam,Cotesta:2020qhw}. Synthetic population catalogs often contain millions of merger events, so SNR calculations for the entire population can sometimes be impractical, requiring the evaluation of a detection probability for each source in the catalog (see e.g.~\cite{Dominik:2014yma,Chen:2017wpg}). Secondly, while GW-related open-source software and documentation are available (e.g. through the Gravitational Wave Open Science Center~\cite{Abbott:2019ebz}), SNR calculations often involve a learning curve for traditional astronomers. One of our goals is to provide them with a fast and accessible tool to compute SNRs, which may be useful for population studies and to rapidly answer questions of interest to astronomers, such as the GW detectability of potential electromagnetic wave sources.

One possibility to speed up these calculation is to interpolate SNR look-up tables. For example, Ref.~\cite{Dominik:2014yma} interpolated SNR tables in the two-dimensional plane $(M_{\rm tot},\,q)$, where $M_{\rm tot}=(m_1+m_2)(1+z)$ and $q=m_1/m_2\geq 1$ are the detector-frame total mass and mass ratio of a binary with component source-frame masses $(m_1,\,m_2)$ merging at redshift $z$. Traditional interpolation methods fail as the dimension of the space of input parameters to be interpolated increases, so this strategy is not easily extended to spinning, precessing binaries.

Deep learning (DL) and neural networks (NNs) have found applications in various GW data analysis tasks, including detection~\cite{George:2017pmj,Gabbard:2017lja,Colgan:2019lyo,Vajente:2019ycy,Dreissigacker:2019edy}, parameter estimation~\cite{Gabbard:2019rde,Chua:2019wwt,Green:2020hst}, waveform modeling~\cite{Setyawati:2019xzw} and population inference~\cite{Wong:2020jdt}.  Simpler but time-consuming tasks which are well-suited to DL (such as multidimensional function interpolation) received relatively less attention from the GW community. We demonstrate that a simple NN can interpolate SNRs produced with different waveform models and detector noise curves with exceptional efficiency. The simple NN used for our proof-of-principle study is very flexible, and it can be trained to interpolate other multidimensional functions with minimal changes.  The goal of this paper is to encourage the GW community to leverage the power of NNs not only on ``classic'' hard problems, but also in relatively simpler tasks which could represent computational bottlenecks in certain applications.  Our source code is publicly available online~\cite{NeuralSNR}.

The paper is organized as follows.
In Sec.\ref{sec:SNR} we describe the SNR calculations needed to produce our training dataset.
In Sec.\ref{sec:Network} we give an overview of our NN architecture, the training procedure, and its performance.
In Sec.\ref{sec:conclusions} we discuss possible extensions of this work.

\begin{figure}[t]
\includegraphics[width=\columnwidth]{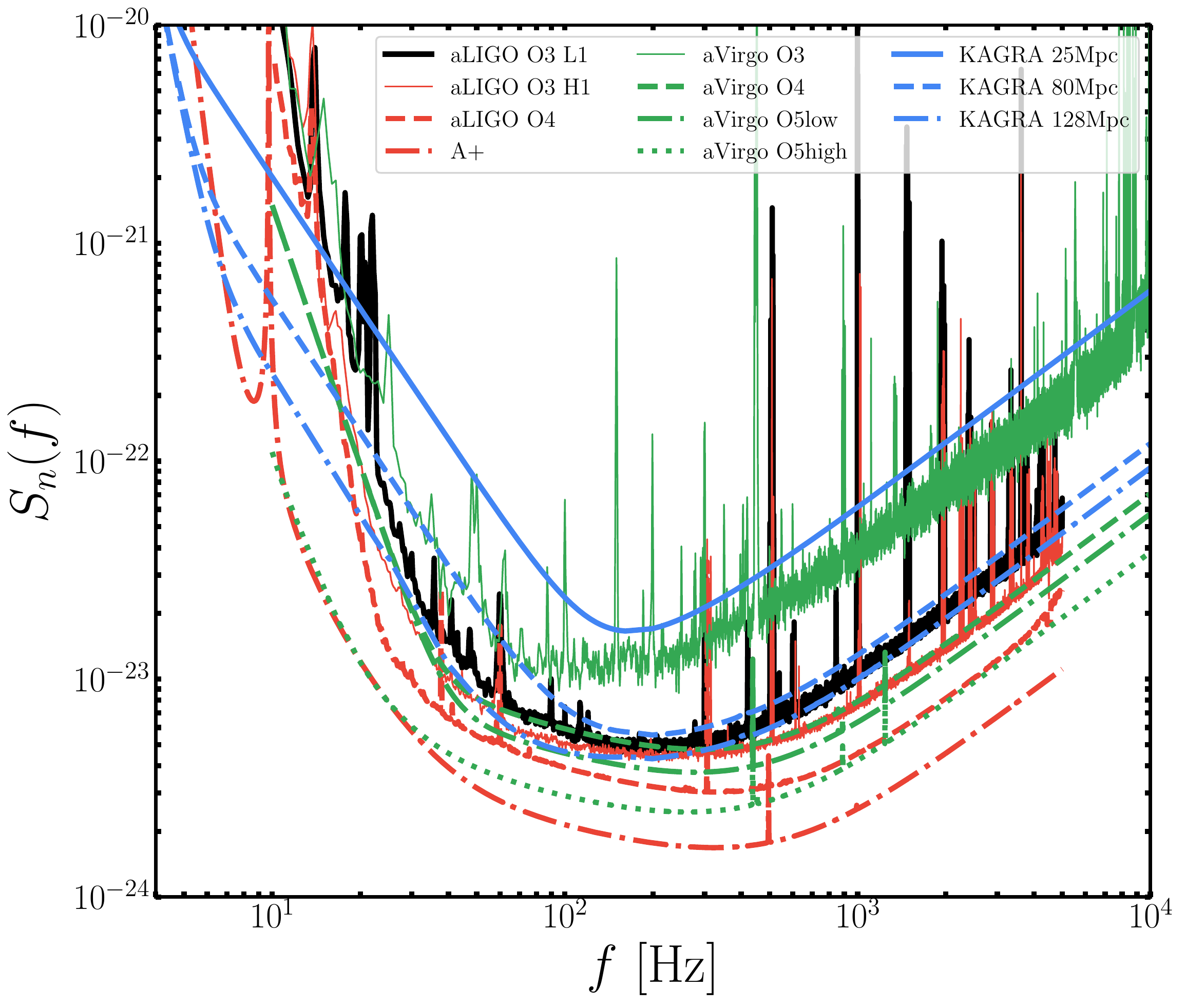}
\caption{Noise power spectral densities used for SNR calculations. These include the O3 sensitivities of LIGO Livingston (aLIGO O3 L1, black), Hanford (aLIGO O3 H1, red) and Virgo (aVirgo O3, green); the projected O4 sensitivities for LIGO (aLIGO O4, red dashed), and Virgo (aVirgo O4, green dashed); the planned A+ upgrade for LIGO (red dash-dotted) as well as pessimistic and optimistic upgrade plans for Virgo (aVirgo O5low, green dash-dotted; aVirgo O5high, green dotted); and three different KAGRA designs labeled by their horizon distance (KAGRA 25\,Mpc, blue; 80\,Mpc, dashed blue; and 128\,Mpc, dash-dotted blue)~\cite{Aasi:2013wya,noisecurve}.}
\label{fig:noiseO4}
\end{figure}

\begin{figure*}[t]
\includegraphics[width=0.95\textwidth]{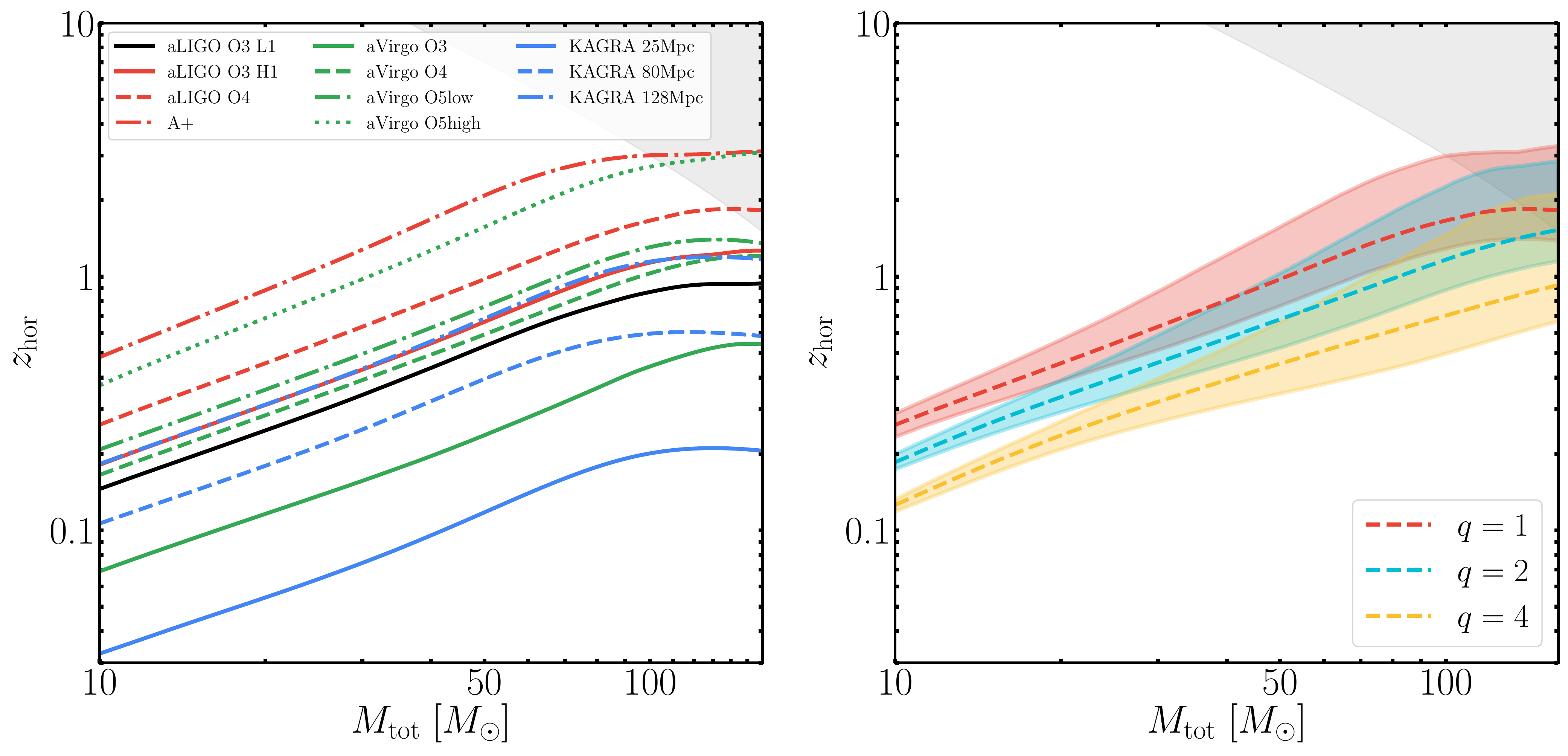}
\caption{Left panel: single-detector horizon redshift $z_{\rm hor}$ as a function of the detector-frame total mass $M_{\rm tot}$ for nonspinning, equal-mass ($q=1$) binaries computed using the {\sc SEOBNRv4} waveform model. Linestyles are the same as in Fig.~\ref{fig:noiseO4}. Right panel: dashed lines show the single-detector aLIGO O4 horizon redshift for a nonspinning binary with $q=1,\,2,\,4$. The shaded regions show the maximum (minimum) horizon redshift for the same value of $(M_{\rm tot},\,q)$, which is achieved for binaries with $\chi_1=\chi_2=1$ ($\chi_1=\chi_2=-1$, respectively) because of the orbital hang up effect.}
\label{fig:hor}
\end{figure*}

\section{SNR calculation}
\label{sec:SNR}

The purpose of this section is to briefly review some basic definitions for the SNR of a GW source, and to spell out the assumptions made in the construction of our training set.
The SNR $\rho(\bm{\theta})$ of a GW source characterized by a vector of parameters $\bm{\theta}$ is given by
\begin{align}
\rho(\bm{\theta}) = \left[ 4\int^{f_{\rm max}}_{f_{\rm min}} \frac{\tilde{h}^*(f;\bm{\theta})\tilde{h}(f;\bm{\theta})}{S_n(f)}df \right]^{1/2}\,,
\label{eq:SNR}
\end{align}
where $\tilde{h}(f;\bm{\theta})$ is the gravitational waveform in the Fourier domain, $f_{\rm min}=5$\,Hz and $f_{\rm max}=4096$\,Hz are the minimum and maximum integration frequencies, and $S_n(f)$ is the one-sided noise power spectral density (PSD).
The evaluation of this integral can be slow, because in some cases the construction of the waveform model is time-consuming: for example, effective-one-body models require the integration of differential equations.
Our goal is to train a NN to approximate $\rho(\bm{\theta})$ using a set of look-up tables for $\rho$ in the multidimensional parameter space spanned by the parameters $\bm{\theta}$.
Once a NN is trained (and checked to be accurate enough), the computational cost of a single SNR evaluation is fixed, and typically lower than the evaluation of Eq.~\eqref{eq:SNR}, because we can bypass the need to generate a waveform and compute the integral.

As a demonstration, we trained NNs using two commonly used waveform models: {\sc IMRPhenomD}~\cite{Khan:2015jqa}
and {\sc SEOBNRv4}~\cite{Bohe:2016gbl}. We used a set of noise PSDs representative of present and near-future GW detectors, as shown in Fig.~\ref{fig:noiseO4}.
Note that we trained all NNs by computing single-detector SNRs for optimally-oriented binaries. The extension to a detector network is trivial.
For nonprecessing systems and ignoring higher harmonics, the sky position coordinates $(\theta, \phi)$, the polarization angle $\psi$ and the orbital inclination $\iota$ can be absorbed in a response function $\Theta(\theta, \phi, \psi, \iota)$ as an overall coefficient, so that the SNR at the detector reads (see e.g.~\cite{Sathyaprakash:2009xs,Schutz:2011tw}):
\begin{align}{\label{eq:rhoTheta}}
    \rho_{\rm obs} = \Theta(\theta, \phi, \psi, \iota) \rho(\bm{\theta}_{\rm int}),
\end{align}
where $\bm{\theta}_{\rm int}=(M_{\rm tot},\,q,\,\chi_1,\,\chi_2)$ is a four-dimensional vector of intrinsic parameters for aligned-spin binaries with dimensionless spin magnitudes $(\chi_1,\,\chi_2)$,
\begin{align}
  \Theta^2(\theta, \phi, \psi, \iota) = \frac{1}{4}F_+^2(\theta,\phi,\psi)(1+\cos^2\iota)^2+F_{\times}^2\cos^2\iota\,,
  \nonumber
\end{align}
and the antenna pattern functions are defined as
\begin{align}
F_+&=\frac{1}{2}\left( 1+\cos^2{\theta} \right) \cos{2\phi}\cos{2\psi}-\cos{\theta}\sin{2\phi}\sin{2\psi} ,\nonumber\\
F_{\times}&=\frac{1}{2}\left( 1+\cos^2{\theta} \right) \cos{2\phi}\sin{2\psi}+\cos{\theta}\sin{2\phi}\cos{2\psi}. \nonumber
\end{align}
In waveform models allowing for precession and higher-order harmonics, such as {\sc IMRPhenomPv3HM}~\cite{Khan:2019kot} and {\sc SEOBNRv4PHM}~\cite{Ossokine:2020kjp}, the inclination dependence of each mode is different, and the factorization in Eq.~\eqref{eq:rhoTheta} breaks down~\cite{Mills:2020thr}. In this case one must extend the parameter space and include the inclination $\iota$ in the training set, but for simplicity we neglect this complication.

We trained our NN to compute optimal SNRs for nonprecessing black-hole binaries located at a fiducial luminosity distance $D_L=100$\,Mpc. The SNR $\rho \propto 1/ D_L(z)$, so we can translate this information into a ``horizon redshift'' $z_{\rm hor}$ within which the binary woud be detectable by a single detector with $\rho(\bm{\theta}_{\rm int})=8$.

The left panel of Fig.~\ref{fig:hor} shows $z_{\rm hor}$ as a function of $M_{\rm tot}$ for an equal-mass, nonspinning binary and for all the noise curves considered in Fig.~\ref{fig:noiseO4}. In the right panel of Fig.~\ref{fig:hor}, where we focus for concreteness on the aLIGO O4 noise, we show how $z_{\rm hor}$ varies as a function of $M_{\rm tot}$, $q$, $\chi_{1}$ and $\chi_{2}$. For a binary of fixed mass, the emitted radiation (and therefore the horizon redshift) decreases as $q$ increases~\cite{Berti:2007fi}. The shaded regions for three selected values of $q=1,\,2,\,4$ show that the redshift horizon can be sensibly larger (smaller) for binaries having both spins aligned (antialigned) with the orbital angular momentum, because of the ``orbital hang up'' effect~\cite{Campanelli:2006uy}.

The computational cost of Eq.~\eqref{eq:SNR} scales with the number of samples needed to achieve a given accuracy in frequency space. The number of samples increases for low-mass events (such as binary neutron stars), which correspond to longer signals. Here we focus on binary black holes with intrinsic parameters in the following ranges: $m_1\in [5,200]\,M_{\odot}$, $q\in [1,\,10^{1.5}]$, $\chi_{1}\in [-1,1]$, and $\chi_{2}\in [-1,1]$.
For each waveform model and sensitivity curve, we create a training set consisting of $8\times 10^5$ training samples, $10^5$ validation samples and $10^5$ test samples with Latin hypercube sampling~\cite{doi:10.1080/00401706.1979.10489755}.
Note that the waveform models we used to generate the training data set come with their own systematic uncertainties, especially in the region where $q\gtrsim 10$.
All SNRs were computed using \texttt{pycbc}~\cite{alex_nitz_2020_3697109}.

\section{Network training and accuracy}
\label{sec:Network}

Here we briefly review a common NN used in this work, limiting our discussion to supervised learning. This section is not meant to be exhaustive, and we omit nonessential technical details (such as the use of back-propagation to speed up the computation): see e.g. Ref.~\cite{Goodfellow-et-al-2016} for a more comprehensive introduction to NNs.

We have a strictly forward problem: learning a function which relates 4 inputs to 1 output. Feed-forward NNs are known as universal approximators~\cite{HORNIK1991251,NIPS2017_7203,2017arXiv170802691H},
i.e. they can approximate any continuous function on compact subsets of $\mathbb{R}^n$ with a finite number of neurons, given mild assumptions on the activation function. Fully-connected multilayer perceptrons (MLPs), which belong to the class of feed-forward NNs, can easily accommodate our task.

\begin{figure}[t]
\includegraphics[width=0.9\columnwidth]{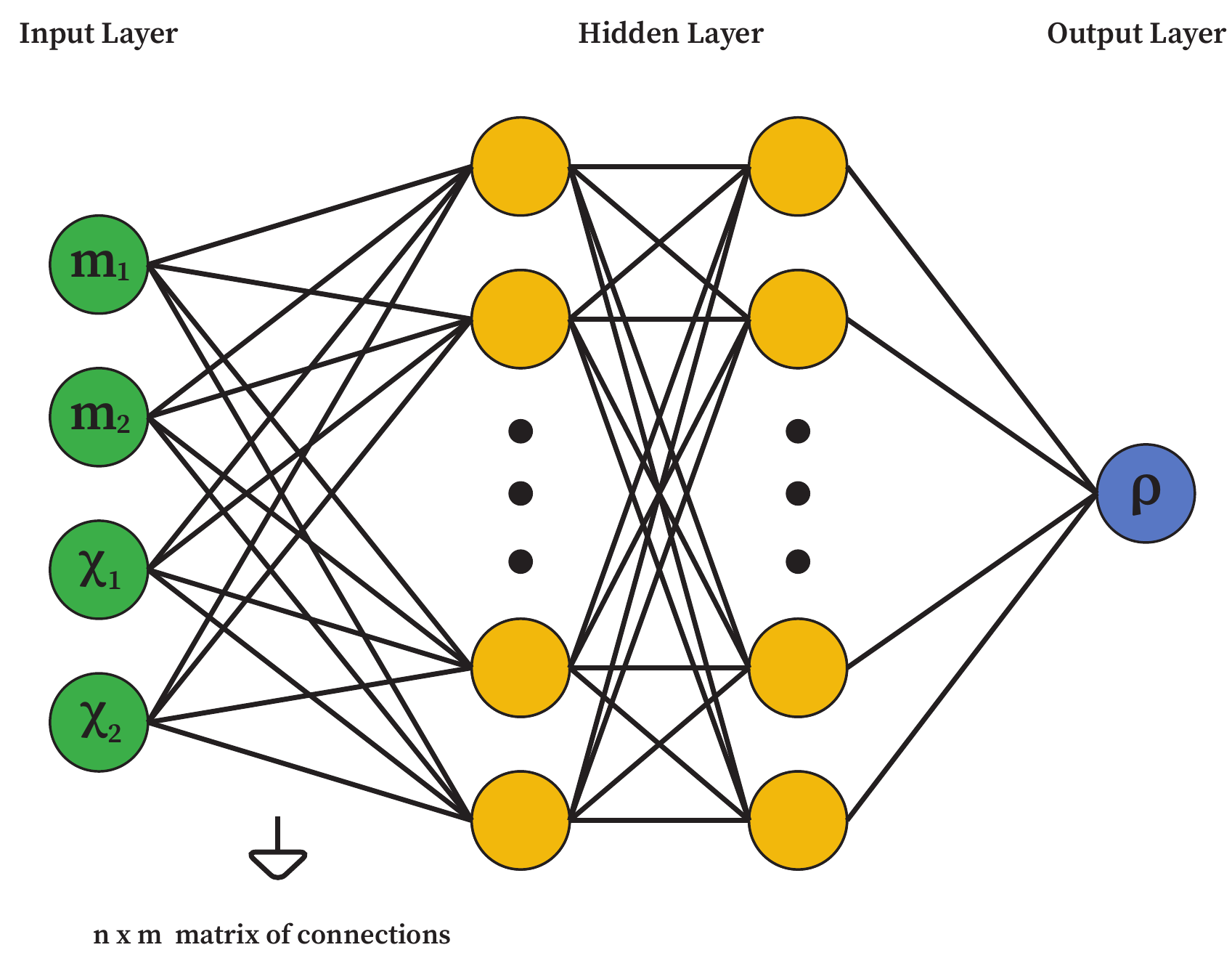}
\caption{Schematic illustration of the architecture of the feed-forward MLP as used in this work. Given the four intrinsic parameters $(m_1,\,m_2,\,\chi_1,\,\chi_2)$, the network is used to interpolate the resulting optimal SNR for a given waveform model and noise power spectral density.}
\label{fig:MLP}
\end{figure}

The architecture of a MLP is shown in Fig.~\ref{fig:MLP}.
The fundamental unit of a MLP is a neuron, usually associated with a scalar quantity $x$ representing its property and parameterized by a bias $b$.
Neurons are organized in layers. A layer with $M$ neurons can then be represented as a vector $\bm{x}$, and the neuron biases are also represented as a vector $\bm{b}$, both vectors having dimension $M$.
A series of layers forms a NN.
Two layers of sizes $M$ and $N$, respectively, can be connected through an $M\times N$ ``weight matrix'' $\bm{W}$,
with matrix elements $W_{mn}$ denoting the connection between neuron $m$ in the first (input) layer and neuron $n$ in the second (output) layer.
To propagate the information from the input layer to the output layer, the vector stored in the output layer will be $\bm{h} = g(\bm{W}^T\bm{x}+\bm{b})$, where $g$ is an element-wise activation function which applies to every element of a vector. Some commonly used activation functions include Rectified Linear Units (ReLUs)~\cite{relu}, sigmoid functions, or hyperbolic tangents; here we use the {\sc swish} activation function~\cite{2017arXiv171005941R}.

By stacking many layers we can form a NN. The input (output) layers interface with the user's input (output), and intermediate layers are called ``hidden''. The number of neurons in each hidden layer can in principle be different, but in practice it is often constant, whereas the number of neurons in the input and output layers must match the dimensionality of the input and output, respectively. Properties such as the number of neurons per layer, the number of layers, and the chosen activation function(s) are called ``hyperparameters''.

The weights and biases (collectively called the ``parameters'') of a NN are initially undetermined. To approximate the output we must train the NN by feeding it training data, and tuning the network parameters until a desired requirement is achieved. Training is often accomplished by minimizing a ``loss function'' through gradient descent methods. Here we adopt the commonly used mean-square error (MSE) loss function $[\sum_{i}^{n}(y_i-\hat{y}_i)^2]/n$, where $\bm{y}$ represents the training data, $\hat{\bm{y}}$ is the network output, and the sum is over the training set.

The main hyperparameters affecting the accuracy and speed of the NN are the number of hidden layers and the number of neurons per hidden layer.  We found 3 hidden layers and 512 neurons to be sufficient for our purpose. Since a NN typically has more parameters than the target function, it could'''overfit'' the function, introducing inaccuracy when we use the NN to evaluate the function at locations not included in the training set.  To avoid overfitting, we compute the MSE loss for a training set and subsequently for a validation set in each epoch, where an epoch represents one loop over the data sets.  The network state such that the validation loss is minimal is our final product.  We trained our NN for 3000 epochs on an Nvidia K80 GPU for $\sim 15$ hours using \texttt{pytorch}~\cite{paszke2017automatic}. The set of NNs trained in this way, as well as the source code used to train them, is publicly available online~\cite{NeuralSNR}.

\begin{figure}[t]
\includegraphics[width=\columnwidth]{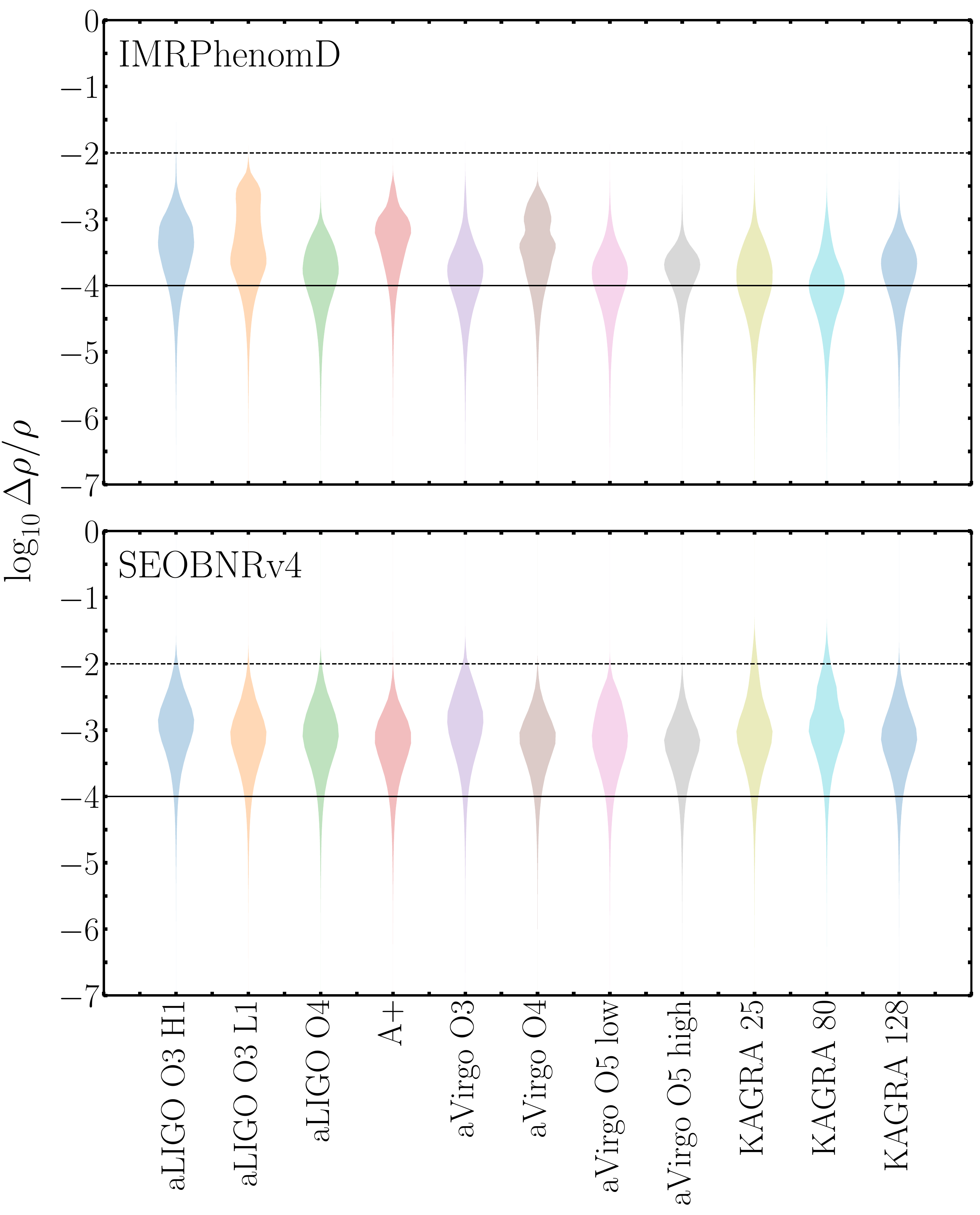}
\caption{Relative error in the computed SNR for the {\sc IMRPhenomD} (top) and {\sc SEOBNRv4} (bottom) waveform models.}
\label{fig:FractionalError}
\end{figure}

To test the accuracy of the NN, we compare SNRs computed for $10^5$ testing samples using the NN against direct integration of Eq.~\eqref{eq:SNR}.
In Fig.\ref{fig:FractionalError} we plot the resulting distribution of fractional errors for the {\sc IMRPhenomD} (top) and {\sc SEOBNRv4} (bottom) waveform models, and for different noise power spectral densities.
Errors typically decrease mildly as the detector sensitivity improves, but this is only a mild effect. If we take the median fractional error for each detector, and then we average over the medians, we get typical fractional errors below $10^{-3}$: $2.22\times10^{-4}$ for {\sc IMRPhenomD}, and $8.17\times10^{-4}$ for {\sc SEOBNRv4}. This is not surprising, since {\sc IMRPhenomD} is a relatively simpler waveform model than {\sc SEOBNRv4} (for example, the {\sc IMRPhenomD} inspiral waveform is a simple TaylorF2 model~\cite{Dudi:2018jzn}).

The main advantage of the NN is the speed gain relative to a numerical evaluation of SNR integrals.
Generating $10^6$ training data sets takes roughly $8\,(5000)$ CPU hours for {\sc IMRPhenomD} ({\sc SEOBNRv4}).
The speed-up is more significant for low-mass, large-$q$ binaries, which have more GW cycles in band.
For example, the generation of an {\sc SEOBNRv4} waveform for a system with $(m_1,\,q)=(200\,M_{\odot},\,1)$ takes ${\sim}100$~ms and the NN speed-up is ``only'' a factor $\sim 1.7\times 10^3$, while for $(m_1,\,q)=(5\,M_{\odot},\,10^{1.5})$ it takes ${\sim}2$~min and the corresponding speed-up is a factor of $\sim 2\times 10^6$.
The trained NN has a fixed number of neurons and connections, and therefore the evaluation time is approximately constant across all input configurations and waveforms.
Our specific NN takes around $1$ minute to evaluate $10^6$ SNRs with 4 CPU cores.
This corresponds to an average speed-up by a factor $120\,(7.5\times10^4)$ for {\sc IMRPhenomD} ({\sc SEOBNRv4}).
Another perk of using a NN is memory usage.
Traditional methods (such as linear interpolation) require the training data to be loaded into memory during runtime, while a NN can be used as a standalone function after training.
Each training data set used in this study uses around 100\,MB of memory, while our network uses about 5\,MB.

\section{Conclusions}
\label{sec:conclusions}

We demonstrated that a simple MLP can significantly speed up SNR calculations with minimal compromises in terms of accuracy. The networks trained in this study should only be used for back-of-the-envelope calculations and prototyping purposes, but they can approximate other multidimensional functions which may not be efficiently fitted by traditional methods, such as polynomial interpolation or splines.

Given suitable training data, this approach can easily be extend to other functions, such as the event rate distribution produced by a population synthesis simulation, or the calculation of the Fisher matrix elements of a GW source.  These and other applications will be explored in future research.

While this paper was being completed, we learned about related work using machine-learning to estimate selection effects in GW observations~\cite{Gerosa:2020pgy}. Their calculation is complementary to ours as it focuses on LIGO/Virgo detectability, rather than on the interpolation of the SNR as a function of the intrinsic parameters of the binary.

\acknowledgments
We thank Davide Gerosa for useful comments on the manuscript. E.~Berti and K.W.K.~Wong are supported by NSF Grants No. PHY-1912550 and AST-1841358, NASA ATP Grants No. 17-ATP17-0225 and 19-ATP19-0051, and NSF-XSEDE Grant No. PHY-090003. This research project was conducted using computational resources at the Maryland Advanced Research Computing Center (MARCC). The authors would like to acknowledge networking support by the GWverse COST Action CA16104, ``Black holes, gravitational waves and fundamental physics.'' K.K.Y.Ng is a member of the LIGO Laboratory. LIGO was constructed by the California Institute of Technology and Massachusetts Institute of Technology with funding from the National Science Foundation and operates under cooperative agreement PHY-1764464.

\bibliography{SNRTable}

\end{document}